**Title**: Automated bird sound recognition in realistic settings

**Authors**

Timos Papadopoulos

Department of Zoology, University of Oxford – ISVR, University of Southampton (current affiliation)

Stephen J. Roberts

Department of Engineering Science, University of Oxford

Katherine J. Willis

Oxford Long-term Ecology Laboratory, Department of Zoology, University of Oxford

Corresponding author: Timos Papadopoulos, timospj@gmail.com

## Abstract

- We evaluated the effectiveness of an automated bird sound identification system in a situation that emulates a realistic, typical application. We trained classification algorithms on a crowd-sourced collection of bird audio recording data and restricted our training methods to be completely free of manual intervention. The approach is hence directly applicable to the analysis of multiple species collections, with labelling provided by crowd-sourced collection. We evaluated the performance of the bird sound recognition system on a realistic number of candidate classes, based upon typical numbers that would be encountered in real conditions.
- Methods: We used a threshold selection method to separate clean bird sound from silence in the crowd-sourced recordings of the training dataset. Test data were obtained from hand-curated recordings and chosen to correspond to an application scenario where the end user selects an excerpt of clean bird sound and presents it, with no extra information, to the identification system. We investigated the use of two canonical classification methods, chosen due to their widespread use and ease of interpretation, namely a *k* Nearest Neighbour (*k*NN) classifier with histogram-based features and a Support Vector Machine (SVM) with time-summarisation features. We further investigated the use of a certainty measure, derived from the output probabilities of the classifiers, to enhance the interpretability and reliability of the class decisions.
- Results: Our results demonstrate that both identification methods achieved similar performance, but we argue that the use of the *k* Nearest Neighbour classifier offers somewhat more flexibility. Furthermore, we show that employing an outcome certainty measure provides a valuable and consistent indicator of the reliability of classification results.

- Wider implications: Our use of generic training data and our investigation of probabilistic classification methodologies that can flexibly address the variable number of candidate species/classes that are expected to be encountered in the field, directly contribute to the development of a practical bird sound identification system with potentially global application. Further, we show that certainty measures associated with identification outcomes can significantly contribute to the practical usability of the overall system.

**Keywords**: Automatic bird sound recognition, Crowd-sourced training data, Machine Learning, Real-world performance evaluation, Flexible geo-temporal species selection, Probabilistic uncertainty measure, Classification rejection option

## Introduction

Over the last decade rapid advances in the sensing capabilities, data storage, network connectivity and computation power of mobile devices have occurred. This has been recognized as a unique opportunity for the deployment of biological recording systems to detect and identify biodiversity using mobile device applications (August et al. 2015). The widespread user base of such devices, as well as the ability to geo-tag & time-stamp observations and to exchange data makes the intelligent gathering of biodiversity information at a massive scale a real possibility. The potential of such methods to underpin the design of data-driven species population models, ecosystem sustainability evaluation and biodiversity protection strategies has been realised in publicly available, either automatic or crowd-sourced, identification applications for birds (ChirpOMatic, Warblr, BirdSongID, Merlin Bird Photo ID)[1], bats (BatMobile, Echo Meter Touch Bat Detector)[2], cetaceans (Automatic Whale Detector)[3], insects (Cicada Hunt)[4] and plants (Plantifier, NatureGate, Leafsnap, Plantsnap, Wild Flower Id)[5].

The use of such approaches to enable automatic bird sound detection and identification started receiving research attention some two decades ago (see e.g. Anderson et al. 1996). In recent years there have been a number of studies carried out to investigate the effectiveness of different audio feature extraction methods and classification algorithms for bird sound identification (see for example Somervuo et al. 2006; Brandes 2008a,b; Trifa et al. 2008; Acevedo et al. 2009; Kirschel et al. 2009; Lakshminarayanan et al. 2009; Farina et al. 2011; Towsey et al. 2013; Wimmer et al. 2013; Stowell & Plumbley 2014b). Identification methods have also been used in the context of audio identification of vocalisations from species, such as bats (see e.g. Walters et al. 2012; Zamora-Gutierrez et al. 2016). More recently, the development of new recognition methods is mainly carried and presented as part of online competitions[6] (see Stowell et al. 2016; Xie et al. 2018 and references therein). The method of choice in these more recent developments is predominantly that of deep convolutional neural networks operating on 2D spectrograms treated as images. Our work does not aim to investigate

---

[1] See itunes.apple.com/gb/app/chirpomatic-uk-automatic-birdsong/id972765162, warblr.net, isoperla.co.uk/BirdSongId.html and merlin.allaboutbirds.org/photo-id/

[2] See batmobile.blogs.ilrt.org and itunes.apple.com/gb/app/echo-meter-touch-bat-detector/id693958125

[3] See itunes.apple.com/us/app/whale-alert-reducing-ship/id911035973

[4] See newforestcicada.info

[5] See itunes.apple.com/us/app/plantifier/id524938919, www.luontoportti.com/suomi, leafsnap.com, plantsnapp.com and itunes.apple.com/gb/app/wild-flower-id-automatic-recognition/id671592368

[6] See www.kaggle.com/c/multilabel-bird-species-classification-nips2013, www.kaggle.com/c/mlsp-2013-birds, www.kaggle.com/c/the-icml-2013-bird-challenge,www.imageclef.org/lifeclef/2016/bird and www.imageclef.org/lifeclef/2017/bird

'best classification' in the narrow machine learning sense. We consider instead the pipeline for creation of detection algorithms, focusing on the role of data for both training and testing as well as issues of labelling. More specifically, the work we present here addresses the following points that we consider important for the development of practical systems for real-world application.

First, with the exception of Lopes et al. 2011, Chou & Ko 2011, Stowell & Plumbley 2014a,b and the most recent online BirdClef competitions, most of the research undertaken to date presents automated classification results associated with approximately 5 to 20 species of birds. This number of candidate classes/species is far below the number of vocalizing avian species that are likely to be encountered in the field. For example, data from the BirdTrack project[7] for the region of Oxford, UK (an area of nine 10-km squares around the city of Oxford) collected by the British Trust for Ornithology list a total of approximately 240 species in an annual cycle[8] with the number of occurring species per week ranging from approximately 80 to 140 (with a mean of 110). An automated classification system deployed in the field would therefore have to operate on a considerably larger set of species than typically investigated in most studies so far and probably around 50-100 classes/species in a typical deployment scenario.

Second, for a species identification method to be scalable to global application, the classifier's training method needs to be applicable to as many of the total number of vocalizing avian species -namely some 10000 species that are found worldwide. Manually assembling training data for such a number of classes and, more importantly, relying on a few individual experts for segmenting and annotating adequate volumes of training data, is thus a major undertaking and not practicable for most regions of the world. This fact renders inapplicable a large number of classification methods presented in the literature so far, especially those that rely on manual segmentation to the syllable or phrase level (see e.g. Anderson et al. 1996; Chen & Maher 2006; Vilches et al. 2006; Lee et al. 2008; Trifa et al. 2008; Acevedo et al. 2009; Vallejo & Taylor 2009; Wei & Blumstein 2011; Aide et al. 2013; Tsai et al. 2014; Tsai & Xue 2014). On the other hand, a publically available database of bird sounds exists, with the sounds both recorded and annotated in a crowd-sourced manner -albeit not at the syllable or phrase level- and provides nearly worldwide coverage of bird species data which is constantly enriched. This database is the xeno-canto project (xeno-canto.org). The identification methods we consider in this paper rely on data obtained from the xeno-canto data source in a manner that is directly and automatically repeatable for any selection of bird species from any region of the world.

[7] BirdTrack is a partnership scheme between the British Trust for Ornithology, Royal Society for the Protection of Birds, Birdwatch Ireland, Scottish Ornithologists' Club and Welsh Ornithological Society.

[8] There are exactly 248 distinct species binomial names in the complete BirdTrack list. However approximately 10 of them correspond to taxonomical splits which we could not trace with absolute certainty (e.g. non-uniquely defined subspecies or hybrids). Hence the numbers we quote here should be taken as an approximation in their least significant digit.

Third, we find that classification methodologies that allow the straightforward and flexible use of time-of-year and location information for the selection of candidate classes (bird species) on-the-spot, for example the $k$ nearest neighbours ($k$NN), have received virtually no attention in the existing literature. Falling under the broader category of 'Instance-based learning' (IBL) classifiers (Mitchell 1997), $k$NN classifiers do not generalise to classification rules during training but rather store training instances and defer further action until a test instance appears for classification. Such information can significantly reduce the number of candidate classes from several thousands (worldwide) down to a few hundred (e.g. in the U.K.) and even fewer (e.g. 50-100 for a given week and a given geographic location in the U.K.). In turn, this can help avoid the discriminative performance degradation due to exceedingly confusable classes (Gupta et al. 2014), make the per-class balancing of training data easier by considering underrepresented species only when necessary and also make trivially easy the incorporation of more training instances as they become available without the need to retrain or implement incremental learning methods.

The overall aims of this paper were therefore to examine the utility of using simple, but flexible, classification methodologies combined with publicly available crowd-sourced training data to aid automatic species identification of bird song excerpts and to evaluate the expected level of performance of such practical applications.

The contributions of this paper are summarised as follows:

- We describe the 'blueprint' of a practical automated bird sound identification system which is directly applicable worldwide.
- We present classification results that directly indicate the expected level of performance and usability of such a bird species identification system.
- We introduce a probabilistic measure of uncertainty associated with the classification output and discuss how this can be used to increase the reliability of the identification results.

## Methods

### Test and Training Audio Recording Data

We used recordings from the 'Reference Animal Vocalisations' section [9] of the Animal Sound Archive dataset as test data (hereafter denoted as RAV recordings). The recordings included in the RAV collection have been manually annotated by the curator of the archive to contain only bird vocalisations of high audio quality with no other sound sources present in the recording. From the collection we selected bird species that contained at least 10 open access recordings. This resulted in a collection of 4132 recordings from 99 species with duration ranging from 0.35sec to 36sec and median duration of 2.26sec (see Table 2 in the Appendix, included in the supplementary material, for details). In terms of a real-world application, this kind of test data correspond to the case where the user singles out a recorded excerpt of clean bird sound (e.g. recorded on their mobile device and 'chopped' from the beginning to the end of a bird vocalisation) and provides it to the recognition system in order to identify the species. Modern mobile devices with large touch display interfaces make such a process feasible on the field.

[9] http://www.animalsoundarchive.org/RefSys/ProjectDescription.php

We used recordings from the xeno-canto online dataset[10] (hereafter denoted as XC recordings) to train the classifiers we investigated. This database currently contains more than 270000 recordings from around the world with tens of thousands of new recordings added every year. For approximately 9300 species there is at least one recording annotated as dominantly containing the corresponding species (species in the 'foreground'). The mean of 'foreground' recordings per species is approximately 25 and the median is 10; there are more than 4600 species represented by at least 10 'foreground' recordings. By selecting XC recordings given a 'Quality A' (highest) rating and marked as having no other species in the background we obtained a collection of 6182 recordings of duration ranging between 0.73sec and 71min42sec and with median duration of 44.7sec (see supplementary material for details).

Even though the selection of species used for our experiments was largely dictated by the availability of publicly accessible, reliably annotated test instances (as was the case with the RAV dataset), it offers a quite good indication of bird species prevalence in Europe. Taking as an indication of prevalence the number of recordings per species currently available on the xeno-canto database, and with the exception of the Canary Islands Chiffchaff (*Phylloscopus canariensis*) which is not a European species, the remaining 98 species used in our experiments contain the 14 most frequent and 18 of the 20 most frequent species in xeno-canto. Out of the total number of 104100 recordings of the 731 European species in xeno-canto, 50914 recordings, nearly the half, are from the 98 European species of our collection.

[10] http://www.xeno-canto.org/

Pre-processing and feature extraction

The audio features we used were based on standard FFT-based spectrograms. We extracted spectral statistics from separate spectrogram frames (frame-level features) and consequently aggregated these over collections of spectrogram frames to create feature vectors. For the test (RAV) recordings, frame-level features were computed for the whole length of the recording while for the training recordings we applied a frame selection method. This method is a modification of that used in the works of Briggs et al. (2009) and Stowell & Plumbley (2014a) and is described in the appendix (see supplementary material). Four types of frame-level features were considered, all choices from the various types described in Briggs et al. (2009) and Stowell & Plumbley (2014a). These were the (i) mean (denoted here as $f_{mean}$); (ii) standard deviation ($f_{std}$); (iii) mode ($f_{mode}$); and (iv) difference between the mode of two consecutive frames ($\Delta f_{mode}$). The feature vectors that were used for classification were consequently determined by computing binned histograms (for the IBL classifier based on Briggs et al. 2009) and time-summarisation statistics (for the SVM classifier based on Stowell & Plumbley 2014a). For the former case we consider 100x50-bin two-dimensional histograms of the pairs ($f_{mean}$ and $f_{std}$) and ($f_{mode}$ and $\Delta f_{mode}$) frame-level features and 100-bin one-dimensional histograms of the $f_{mode}$ frame-level feature. For the latter case we use the 6-dimensional time-summarisation features described in Stowell & Plumbley (2014a) comprising the 5th, 50th and 95th percentiles of the $f_{mode}$ and the 50th, 75th and 95th percentiles of the $\Delta f_{mode}$ frame-based features (see the appendix for details on the spectrogram computation parameters).

In both cases (histogram binning and time summarisation) the feature vectors used for training were obtained from sequences of 100 frame-level features (from one or more XC recordings) selected by the power threshold method mentioned above. With the chosen spectrogram parameters (listed in the appendix), each such sequence corresponds to between 1sec and 2sec of clean bird sound. The feature vector for each test instance (RAV recordings) was computed by histogram binning or time summarisation over the whole length of the recording.

The total number of frames selected by use of the power threshold method from the XC recordings for each species is listed in Table 2 in the appendix. We balanced the training dataset by subsampling according to the class with the fewer selected frames. As can be seen in Table 2, this is *Emberiza pusilla* for which the selection method returned in 1830 training frames. Following the procedure described above, 18 '1sec' training instances of 100 frames each were randomly selected (without resubstitution) for each species. We also investigated a second selection of species, namely the 72 species for which the power threshold selection method returned at least 20000 frames (again listed in Table 2). In this case, 200 training instances were used, again comprising 100 frames each. The corresponding test dataset (RAV recordings) for the collection of 72 species comprised 3354 RAV recordings (see Table 2).

Classifier and performance evaluation methods

The IBL-type classifier that we investigated is a *k*NN classifier. It draws from the work presented in Briggs et al. (2009) where a nearest neighbour classifier with time-distribution histograms as feature vectors was tested successfully for classification of 6 species. In the present work we modified this method to a *k*NN voting scheme with a tie-breaking rule (rather than a single nearest neighbour methodology as was used in Briggs et al. (2009)). We compared the results of this modified classification scheme against a more recent work (Stowell & Plumbley 2014a) in which a Support Vector Machine (SVM) classifier was used in a setup that again adheres to the practical application requirements outlined in the introduction (i.e. training data obtained from field recordings only annotated at the recording level, with no manual segmentation to the phrase or syllable level).

It is important to note that in both the aforementioned works, the training and test data come from the same dataset, although some care was taken to avoid training and test data sharing an individual bird origin. In an effort to better investigate real-world conditions of application, in our study we used completely separated training and test data.

We used the L1 (Manhattan) distance between histogram feature vectors for the *k*NN classifier (we also investigated the use of the Kullback–Leibler divergence and an approximation to the Hellinger distance as described in Briggs et al. (2009) in small-scale tests and the performance was not influenced considerably). We took the voting score with an addition of a tie-breaking bias (see details in the supplementary material) as the posterior probability of class membership in the case of the *k*NN classifier (Bishop, C. M. 2006 pp. 124-126). For the SVM classifier we used the Matlab interface of the LIBSVM software package[11] with its default settings for multiclass probabilistic classification (Chang & Lin 2011). These settings amount to a radial basis function kernel and a "one-against-one" method for recasting the problem of training a $k$-class multiclass classifier to the training of $k(k\text{-}1)/2$ binary SVM classifiers. We used the estimate provided by LIBSVM for the posterior class-membership probabilities which is based on the methods described in (Wu et al. 2004 and Lin et al. 2007).

A concept that can be employed for the improvement of the performance and the usability elicited from a probabilistic classification scheme, is the separation of the probabilistic inference and classification decision stages and the introduction of a classification rejection region (Bishop, C. M. 2006). It is worth noting that despite the fact that both the $k$NN and SVM methods considered in the present work provide a probabilistic output, this is effectively disregarded in the classification stage with the decision made solely on the ranked list of probabilities and not their actual values. Further to that we note that, to the best of our knowledge, none of the works presented so far on the subject of automated bird sound identification have investigated this possibility.

In a binary classifier (and assuming constant gain for correct classification and loss for misclassification), the expected loss due to a wrong classification decision is related to how close to unity is the assigned class-membership probability (Bishop, C. M. 2006). For the application of this notion to a multiclass setup, we use here the *entropy* (Bishop, C. M. 2006) of the class-membership probability vectors as the classification rejection criterion. Being a measure of the information content encoded in the probability distribution of a discrete random variable, the entropy value can be used as an indicator of the certainty associated with the classification decision. In simple terms, low values of entropy correspond to class-membership probability distributions that tightly peak in one class or a small number of classes (and are hence associated with higher certainty) whereas high levels of entropy correspond to distributions that are closer to the uniform distribution.

[11] www.csie.ntu.edu.tw/~cjlin/libsvm/

We used the Receiver Operating Characteristic (ROC) in a one-class-versus-all setup for the evaluation of the classification performance. In a binary probabilistic classifier, the ROC curve traces the points with coordinates equal to the achieved true positive and false positive rates as the value of the probability threshold discriminating the positive from the negative class ranges from 0 to 1 (Fawcett, 2006). The constructed Area Under Curve (AUC) metric ranges from 1 (absolutely correct assignment of instances to classes) to 0 (assignment of all instances to the opposite class) with a value of AUC equal to 0.5 corresponding to a classification result equivalent to chance assignment of test instances to classes. In the multiclass setup considered here, the AUC was computed taking each class in turn as the positive with the remaining classes taken as negative. Using a binary classification example of a highly non-balanced dataset, Davis & Goadrich (2006) show that the area under the Precision-Recall curve can be a more informative metric than that of the ROC curve. We include results of that metric again in a one-class-versus-all setup.

The AUC-ROC metric offers a method for the comparative evaluation of different classifiers' performance that is robust in the case of non-balanced test datasets and which is arguably superior to accuracy-based evaluation methods even when a balanced test dataset is used (Huang and Ling, 2005). However, its intuitive interpretation (namely, the probability of a randomly chosen negative test instance being ranked by the classifier lower than a randomly chosen positive test instance) does not lend itself to a direct gauge of the practical effectiveness of an identification system such as the one investigated here.

If we consider the use case where the classifier returns an ordered list (of a chosen length $N$) of most likely species, the most easily interpretable measure of its effectiveness would be how often the correct species is within the returned list as a function of its length; a measure that we call *accuracy@N* and which we evaluate in the results section on a balanced test dataset.

Finally, a measure that is more widely established in the topic of information retrieval, and which unlike raw accuracy takes into account not only whether the correct class is within the returned ordered list of length $N$ but also how high it is on that list, is the *mean average precision at N* (MAP@$N$) (Manning et al. 2009). In the case where there is only one relevant retrieval option (as our classification setup) this metric becomes equivalent with the *mean reciprocal rank at N* (MRR@$N$) which is used in the results section below. The MRR@$N$ metric ranges from 0 (in the case where the correct class is not within the $N$ returned classes for any of the test instances) to 1 (when all test instances return the correct class ranked as highest).

**Results**

Figure 1 plots the summary statistics of the classification results (median, interquartile range and whole range of the vector of AUC numbers for the ROC curve for each class against the rest, as well as the mean of the per-class AUC-ROC vector weighted by the number of test instances in each class). Figure 2 plots the same statistics but this time for the AUC of the Precision-Recall curve. The same results are given in tabular from in Table 1. Furthermore, while noting that this is not a consistent evaluation metric in a non-balanced test set (see Huang and Ling, 2005), in Table 1 we list for completeness the corresponding accuracy scores.

The weighted mean and the median AUC-ROC for the histogram-based *k*NN classifier are practically constant over different choices of audio features and number of voting neighbours and marginally higher for the 72 species than the 99 species case. The AUC-ROC results for the SVM classifier with time-summarised audio features are slightly lower in all cases. The accuracy results show an increase in performance with increasing number of voting neighbours. Again, in terms of accuracy, the ($f_{mode}$, $\Delta f_{mode}$) features perform better compared to ($f_{mode}$) and ($f_{mode}$) perform better than ($f_{mean}$, $f_{std}$) . The same characteristics are associated with the AUC of the Precision-Recall results.

Despite the fact that the AUC-ROC results were nearly constant in the *k*NN experiments, there was a clear differentiation in the per-species profile of the performance achieved for different choices of features. For all three features cases ($f_{mean}$ and $f_{std}$ ; $f_{mode}$ ; $f_{mode}$ and $\Delta f_{mode}$), the Pearson correlation between the one-class-versus-rest AUC-ROC vectors obtained with the same features but different numbers of voting neighbours (1, 5, 11, 17 for the 99 species dataset) ranged from a minimum of 0.978 to a maximum of 0.998. For the 72 species dataset and with number of voting neighbours taking the values 1, 5, 11, 21, 51, 101, 201, the Pearson correlation between AUC-ROC vectors obtained with the same features ranged from 0.944 to 0.999 indicating that different species were recognised consistently better with different types of audio features. Contrary to that, keeping the number of voting neighbours constant and changing the features used, resulted in a Pearson correlation between the obtained AUC-ROC vectors that ranged from 0.576 to 0.893 and from 0.586 to 0.863 for the 99 and 72 species datasets respectively.

As can be seen in Table 1, the highest AUC-ROC score is achieved in 10 out of the 12 *k*NN setups for the Common grasshopper warbler (*Locustella naevia*). Garden warbler (*Sylvia borin*) and Hooded crow (*Corvus cornix*) scored the highest AUC-ROC in the remaining two *k*NN cases of Table 1 while Canada goose (*Branta canadensis*) and European nightjar (*Caprimulgus europaeus*) scored highest for the two SVM cases. This level of performance was consistent for these species on a total of 13 parameter setups for the 99 species dataset (three feature cases and 1, 5, 11, 17 voting neighbours, plus the SVM setup) and 22 parameter setups for the 72 species dataset (three feature cases and 1, 5, 11, 21, 51, 101, 201 voting neighbours, plus the SVM setup). A score of more than 0.95 ROC-AUC was obtained 12 and 22 times respectively for the Common grasshopper warbler. The same performance (more than 0.95 AUC-ROC), was obtained 9 times in the 99 species dataset for the Canada goose (that species was not in the 72 species dataset), 4 and 19 times respectively for the Hooded crow, 4 and 14 times respectively for the Garden warbler and 1 and 8 for the European nightjar. Other than the species appearing in Table 1, the same level of performances was also obtained for the Eurasian wren (*Troglodytes troglodytes*) 13 and 18 times respectively and for the Common firecrest (*Regulus ignicapilla*) 12 times in the 99 species dataset (that species was not in the 72 species dataset). We did not find any systematic misclassifications between species pairs in our experiments.

In order to get a more direct measure of the performance, we also obtained the accuracy@*N* metric on a balanced test dataset. We used the same 99 and 72 species collections and the same training procedure with xeno-canto data that was described above. We selected 10 recordings for each species from the Animal Sound Archive dataset as the test set. Test recordings were chosen by minimizing the number of test instances labelled as coming from the same individual (the identifiers of the chosen recordings are provided in the supplementary material). In Figure 3 we plot the percentage of classification results in which the correct species was within the first *N* returned classification outputs, as a function of *N*. In these results we compared the SVM classifier discussed above with one case of the *k*NN classifier, namely the one where one-dimensional histograms of the $f_{mode}$ frame-level feature are used and the number of voting neighbours is set to 17 and 101 for the 99 and 72 species datasets respectively. The other parameter settings of the *k*NN classifier returned very similar results. In the case where only one class was returned by the classifier, the accuracy can be seen in Figure 3 to be equal to 9% and 6% for the *k*NN and SVM classifiers for classification among 99 species and 16% and 10% for classification among 72 species. The corresponding accuracy scores for a returned list of 10 species by the *k*NN classifier was approximately 40% (50% for classification among 72 species) and approximately 30% and 40% respectively for the SVM classifier. Interpreted with the expected user engagement in mind, this identification performance is still in need of improvement. When the number of candidate species is limited to 72 (a number which is at the lower end of the species expected to be encountered in the field) the correct species is expected to be the top result of the best performing *k*NN classification method approximately one in six times. Even when the identification scheme is allowed to provide the 10 most probable results (a list length that is already rather cumbersome to display on a mobile device) the correct species is expected to be within the returned list of the *k*NN classifier only half of the time.

In Figure 4 we plot the MRR@10 obtained for the same training and test datasets and the same classifier configurations as in Figure 3 as a function of the proportion of test instances that are classified using the entropy-based rejection criterion described in the 'Methods' section. The application of the rejection criterion is effected by taking a uniform grid of possible entropy levels ranging from 0 to the maximum level of entropy (equal to ln(99) = 4.59 and ln(72) = 4.28 for the 99 and 72 species cases) with points spaced at a distance of 0.1. For each of these entropy levels, instances associated with higher entropy are rejected prior to the determination of the MRR@10 performance.

The MRR@10 metric when all instances are classified is equal to 0.18 and 0.12 for the $k$NN and SVM classifiers respectively in the 99 species dataset and 0.28 and 0.18 in the 72 species dataset. The performance is consistently rising as the rejection threshold becomes more stringent (i.e. the maximum allowed level of entropy for trusting a classification is reduced). It reaches the maximum MRR score of 1 albeit at very high classification rejection levels (approximately 99%) with the exception of the $k$NN results of the 72 species dataset that show a drop in performance at the 98% classification rejection level. Starting at lower performance levels when all tests samples are accepted for classification, the SVM classifier shows better performance at higher classification rejection rates. It thus appears that the per-class assignment of posterior probabilities by the SVM scheme is more successful than the nearest neighbour scheme. All the aforementioned characteristics appear consistently in the other cases of frame-level features and number of voting neighbours settings of Table 1 (not considering the single nearest neighbour setting for which the entropy-based classification rejection criterion is clearly not applicable). The consistent behaviour displayed in these results suggests that entropy can indeed provide a quite useful classification reliability measure, but evidently, very high performance is only obtained when a very large percentage of tests is not classified

An evaluation of the introduced classification reliability measure which is more directly related to practical application is presented in Figure 5. In that figure we plot again the MRR@10 metric but this time as a function of the entropy threshold level (i.e. the maximum value of entropy over which the classification result for a test sample is not considered reliable and is not included in the evaluation). To account for the fact that the entropy scaling is different for different numbers of candidate classes, we normalise the entropy value by dividing it by the maximum possible value of entropy in each of the 99 and 72 classes cases. Such a normalised value of the reliability measure can be provided to the user of an application together with the list of most likely species returned by the probabilistic classifier. As can be seen in Figure 5, with the exception of the SVM results in the 99-classes experiment, all other cases show a consistent behaviour whereby a marked increase in reliability occurs for values of normalised entropy below 0.6 with values below 0.4 being associated with MRR@10 equal or exceeding a value of approximately 0.7. While there is clearly need for a more detailed investigation, these results suggest that a calibration of the introduced reliability measure which is consistent over different sets of candidate species may well be possible a least for the $k$NN method.

**Discussion – Conclusions**

The results presented in this paper are focused on classification of bird audio recordings among a large number of species (as is realistically required) making use of training data that are presently available for nearly global scalability. The rationale for undertaking this work was driven by a perceived knowledge gap in the expected level of performance of such a practical bird audio identification system. For example, among previous related investigations, the work presented in Lopes et al. (2011) covers audio recordings from 73 species from the Southern Atlantic Brazilian Coast but presents classification results for up to only 20 species. The performance of the methods they investigate is consistently reduced as the number of classes is increased (from 3 up to 20). Chou & Ko (2011) present classification results among 420 species of Japan birds. However, in this study little information is available about the characteristics of their training audio dataset and about the actual degree of train and test data separation in the experiments.

In addition to the SVM-based classification method presented in Stowell & Plumbley (2014a) (and which we partly replicated showing similar performance with the *k*NN method investigated here) the same authors have investigated (Stowell & Plumbley, 2014b) the use in bird audio recognition using the unsupervised feature learning method of Coates, A., & Ng, A. Y. (2012). This approach is applied to various classification experiments of recordings containing vocalisations from approximately 80 to 500 species with very positive results. On the other hand, their feature learning method seems to be closely tied to the particular instance of training data and selection of candidate classes. It is thus questionable if it can be easily applied to an ad-hoc selection of candidate species without significant retraining requirements. The authors of that work also find that their method is possibly demanding in terms of the amount of training data required in order to achieve its best performance.

The IBL classification scheme we investigated in this paper mainly draws from the method proposed in Briggs et al. (2009). The use of higher dimensional histograms with codebook clustering during the training process was also investigated in that work. The performance improvement that was quoted with that approach was rather moderate (increase in accuracy from 88% to 92% for the leading choice of parameters in their experimental setup while at the same time tying the training process to a particular training dataset and selection of candidate classes). On the other hand, we found that the introduction of the $f_{mode}$ frame-level feature (position of maximum frequency) instead of the $f_{mean}$ and $f_{std}$ spectral statistics achieved the same performance in our experimental setup by using only one-dimensional histograms (and hence having reduced requirements in storage size and computation time). We also found that, with the use of a tie-breaking bias addition, the performance of the histogram-based *k*NN classifier is practically constant over the number of voting neighbours. This seems to be in agreement with the finding in Briggs et al. (2009) that taking into account the distance of more distant neighbours in a Bayes risk minimizing classifier formulation gives practically identical results with the single nearest neighbour approach.

Rather than offering gains in performance compared to the single nearest neighbour case, the extension to a *k*NN method (with a probabilistically interpretable output) allows the determination of an uncertainty measure of the classification result as an additional output. It is evident that among classification systems, it is the moderately performing ones (such as many current-day bird audio recognition systems) that can significantly benefit in terms of their practical usability from a consistent uncertainty measure of the output. At the expense of not having a result in many cases, a large number of very likely wrong classification results can be discarded and a smaller number of classification results can be relied upon. In our experimental setup, when all test instances are classified (no rejection option introduced), the combination of histogram-based features with a *k*NN classifier performs better than the SVM method using time-summarised features but this performance comparison is reversed at high classification rejection rates. We cannot yet conclusively determine whether this result is due to the different classifiers or the different types of features and whether it is systematic in different sets of species/classes.

As discussed, the flexibility of an instance-based classifier comes at the cost of increased data storage requirements and computation power in the classification stage. Taking as an example the xeno-canto training data used for the set of 99 bird species and the balancing subsampling of the training dataset in the experiments presented here, the required data storage (making use of efficient storing of the histograms' sparse arrays) ranges from less than 1MByte to approximately 35Mbytes for the different types of audio features used here. The processing time needed for feature computation and *k*NN classification of the whole set of 4132 test instances (total time duration of 198 minutes) in the Matlab programming environment running on a PC with Intel Core i3-4160 CPU@3.6GHz, ranged from approximately 250sec to 500sec (for the different histogram feature and classification parameters considered here). This corresponds to a maximum of 40ms of computation for each second of recorded sound. Evaluating how these requirements would translate in a practical deployment for mobile devices is another important challenge.

The process that we described for the creation of the IBL classifier training instances from xeno-canto recordings is completely free of manual intervention. When combined with (i) the fact that such crowd-sourced data sources offer practically global coverage of bird species recordings and (ii) with the ability to choose the candidate species at run-time with no need for further classifier training, the IBL methodology that we present in this paper offers a blueprint for the development of a globally applicable bird sound identification system on mobile devices which readily provide geo-location information. Our further work plans include larger-scale experiments using more recently established standard datasets (such as the BirdCLEF dataset [12]) for the determination of the bearing of different feature sets and classifier parameters on the system's performance as well as the direct comparison with other classification methods. Our plans also include the use of existing detailed global coverage species distribution information (provided for research purposes by the 'Bird species distribution maps of the world' project[13]) for the scaling of those evaluation experiments to scenarios of application to different geographical regions..

[12] http://www.imageclef.org/lifeclef/2017/bird

[13] BirdLife International and NatureServe (2015) Bird species distribution maps of the world. BirdLife International, Cambridge, UK and NatureServe, Arlington, USA.

In conclusion, in this paper we present evaluation results that directly quantify the expected performance of a practical automated bird sound identification system. We focus on the use case scenario where a mobile device user provides an excerpt of recorded bird sound along with geographical and temporal information provided by the device. The latter is used to select a sufficiently (but not unnecessarily) extensive list of candidate species. In our work, publicly available, crowd-sourced, training data was used in a fashion free from manual preprocessing. We couple this with an appropriately flexible classification scheme to provide a list of most likely species identification results. The current, constantly increasing, collection of bird audio recordings from the xeno-canto dataset allows the described identification scheme to have global application. The results we present compare favourably with previous work adhering to similar application requirements. We evaluate the application of a method for the improved use of the classifier's probabilistic output in refining the classification. There is, however, significant room for improvement in terms of the accuracy of the results presented to user in order for a system such as the one described here to be positively appealing and engaging. To that end, our current work is focused in tuning and optimizing the parameters of the training data selection and audio feature extraction methods in an effort to further improve the identification performance.

**Acknowledgements**

Dr. Timos Papadopoulos is supported by a James Martin Fellowship from the Oxford Martin School, University of Oxford. We are thankful to Dr. Andy Musgrove of the British Trust for Ornithology, Dr. Paul Jepson of the School of Geography and the Environment, University of Oxford and Dr. Karl-Heinz Frommolt of Museum für Naturkunde, Berlin for sharing data and for providing valuable insights and discussions.

## Tables

| Classifier | Features | Voting neighbours | Per-species AUC (ROC) in one class v. all | | | | Species of maximum and minimum AUC (ROC) | | Per-species AUC (Precision-Recall) in one class v. all | | Accuracy (%) |
|---|---|---|---|---|---|---|---|---|---|---|---|
| | | | Min | Max | Weig. Mean | Median | Min | Max | Weig. Mean | Median | |
| Results for 99 species dataset | | | | | | | | | | | |
| *k*NN (histograms) | $f_{mean}$ $f_{std}$ | 1 | 0.44 | 0.98 | 0.77 | 0.80 | lox_cur | loc_nae | 0.086 | 0.024 | 4.53 |
| | | 17 | 0.42 | 0.98 | 0.77 | 0.79 | lox_cur | loc_nae | 0.110 | 0.035 | 6.17 |
| | $f_{mode}$ | 1 | 0.50 | 1.00 | 0.78 | 0.81 | emb_aur | loc_nae | 0.084 | 0.031 | 6.58 |
| | | 17 | 0.49 | 1.00 | 0.78 | 0.81 | chl_chl | loc_nae | 0.102 | 0.033 | 7.41 |
| | $f_{mode}$ $\Delta f_{mode}$ | 1 | 0.46 | 1.00 | 0.79 | 0.81 | tur_pil | loc_nae | 0.089 | 0.034 | 7.74 |
| | | 17 | 0.46 | 1.00 | 0.79 | 0.80 | tur_pil | loc_nae | 0.112 | 0.037 | 8.66 |
| SVM (summ. stats) | $f_{mode}$ $\Delta f_{mode}$ | - | 0.31 | 0.97 | 0.73 | 0.76 | lox_cur | bra_can | 0.091 | 0.024 | 6.15 |
| Results for 72 species dataset | | | | | | | | | | | |
| *k*NN (histograms) | $f_{mean}$ $f_{std}$ | 1 | 0.29 | 0.96 | 0.78 | 0.81 | lox_cur | syl_bor | 0.120 | 0.042 | 8.05 |
| | | 101 | 0.25 | 0.98 | 0.79 | 0.81 | lox_cur | cor_cor | 0.196 | 0.078 | 9.93 |
| | $f_{mode}$ | 1 | 0.62 | 0.99 | 0.80 | 0.82 | jyn_tor | loc_nae | 0.118 | 0.045 | 9.33 |
| | | 101 | 0.61 | 0.99 | 0.82 | 0.84 | den_maj | loc_nae | 0.173 | 0.069 | 11.24 |
| | $f_{mode}$ $\Delta f_{mode}$ | 1 | 0.58 | 0.97 | 0.79 | 0.82 | tur_pil | loc_nae | 0.130 | 0.056 | 13.06 |
| | | 101 | 0.62 | 0.98 | 0.81 | 0.84 | tur_pil | loc_nae | 0.197 | 0.096 | 15.06 |
| SVM (summ. stats) | $f_{mode}$ $\Delta f_{mode}$ | - | 0.23 | 0.98 | 0.75 | 0.77 | lox_cur | cap_eur | 0.143 | 0.036 | 6.89 |

Table 1. Summary of per-species AUC performance metric obtained with the ROC and Precision-Recall curves in a one class v. all setup. Species binomial names in rightmost columns are abbreviated to 3 first letters for the genus and the species. The top 7 rows correspond to the 99 species dataset and the lower 7 rows to the 72 species dataset. In the right-most column we also give the accuracy score (in percent values) for each case.

## Figures

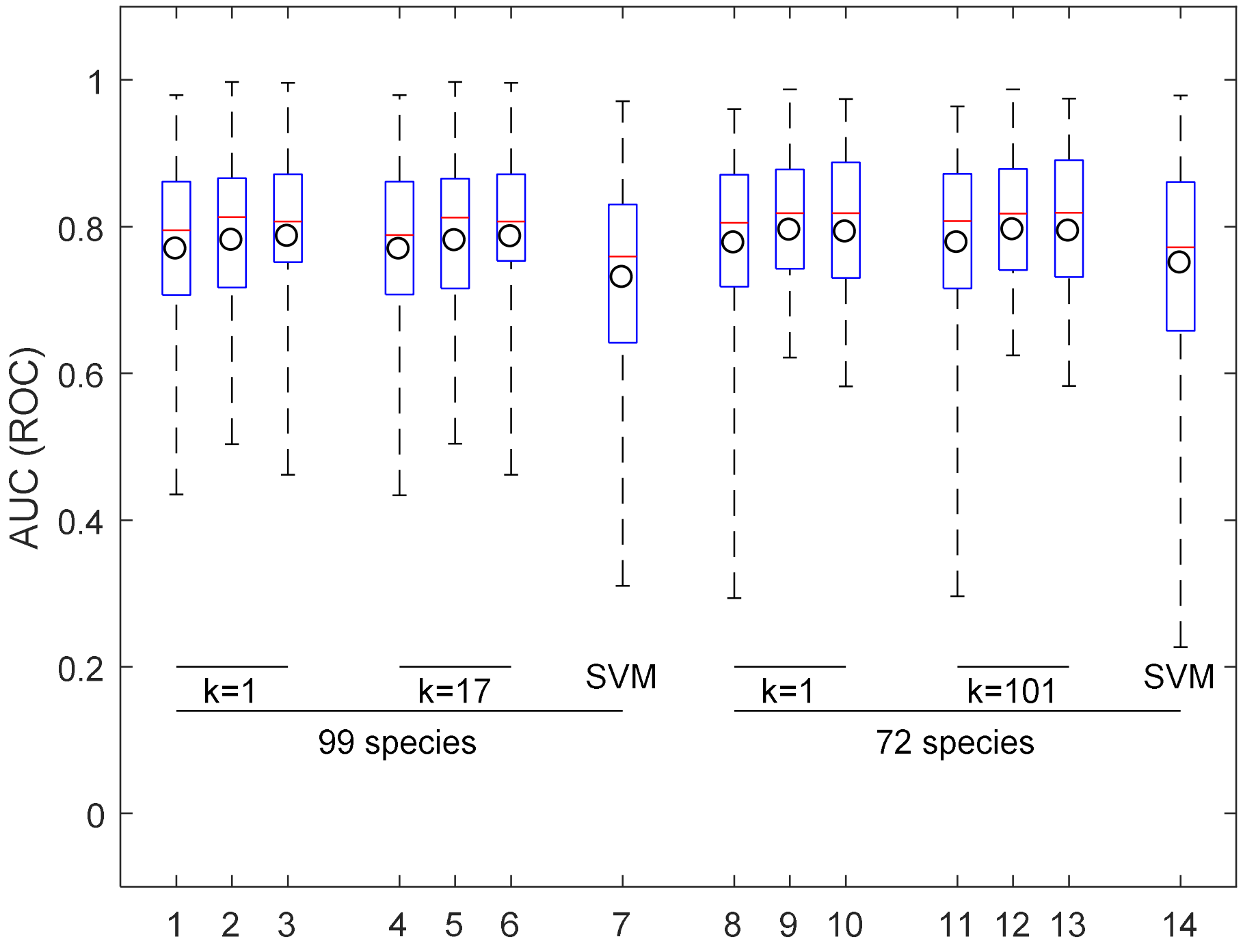

Figure 1: Boxplots are median, interquartile range and whole range of the vector of Area Under the ROC Curve performance for each class against the rest. The black circle is the mean of the per-class AUC vector weighted by the number of test instances in each class. The boxplots appearing in triplets correspond (in order from left to right) to the three cases of frame-level spectral features used in the histogram kNN classifier, namely (i) mean and standard deviation of the frame spectrum, (ii) position of the maximum frequency and (iii) position of maximum frequency and frequency modulation across successive frame couples. For the SVM case we use a 6 dimensional vector of summary statistics.

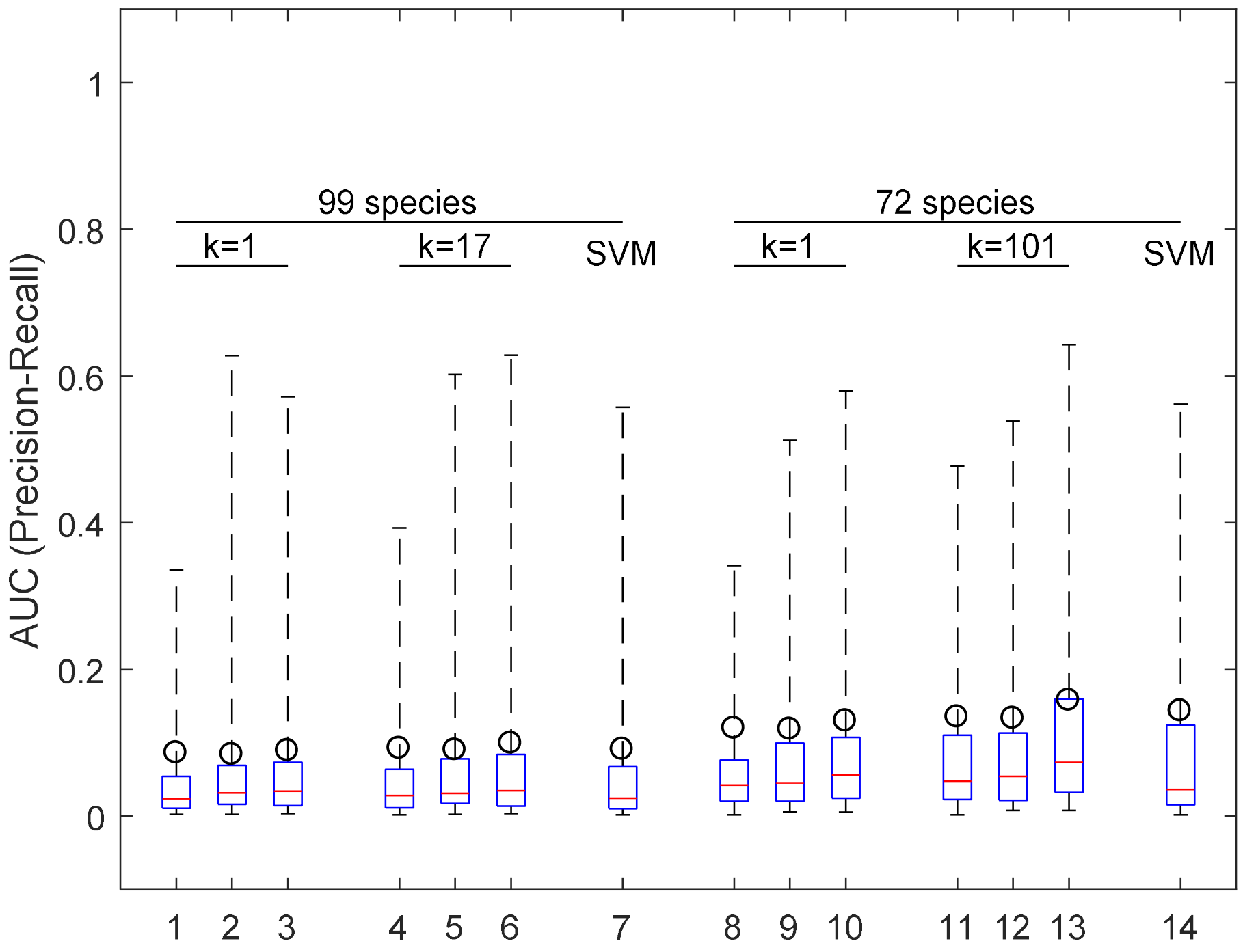

Figure 2: Same as in Figure 1 but for the Area Under the Precision-Recall Curve.

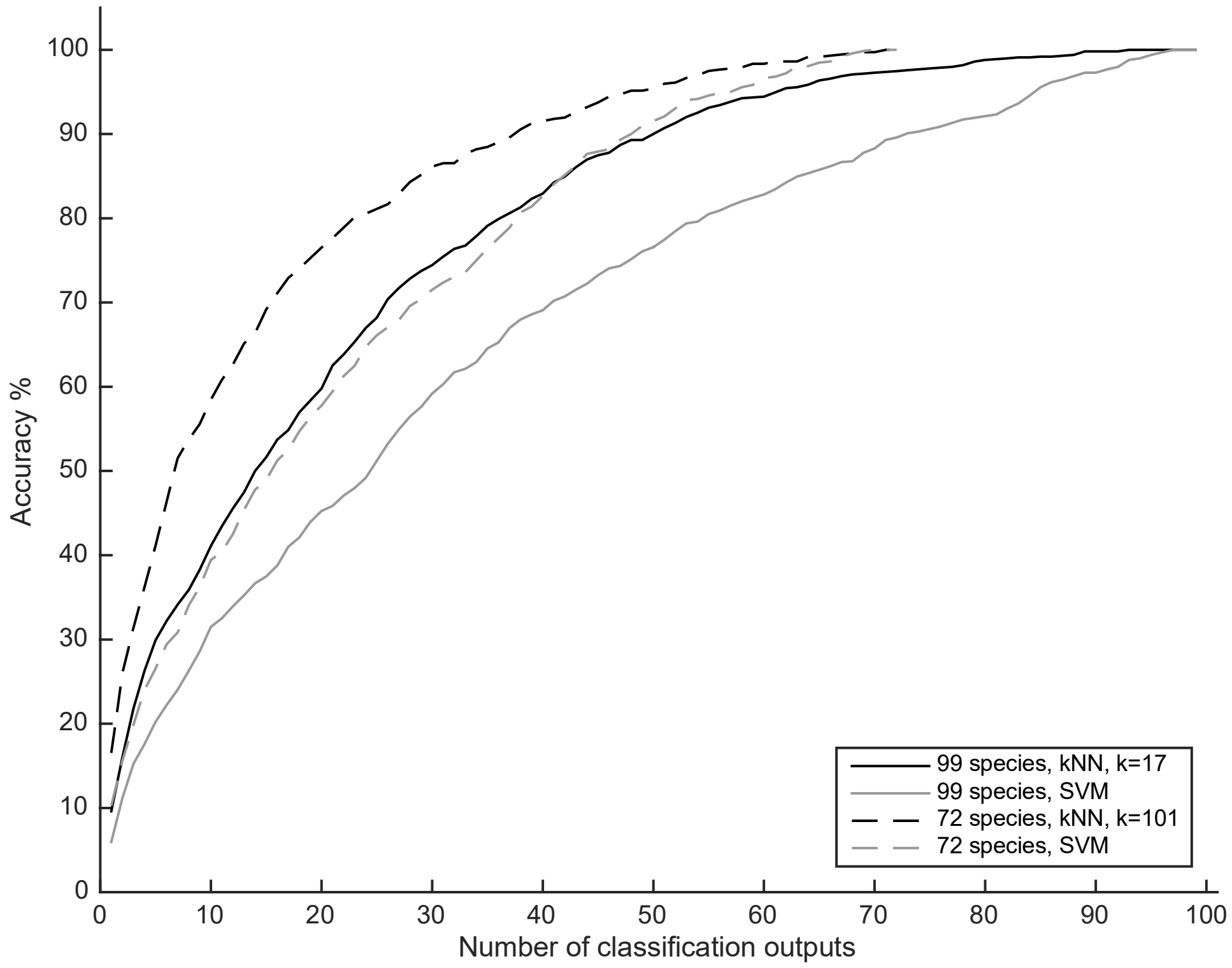

Figure 3: Accuracy in the case where the classifier returns a list of $N$ most probable classes. The plotted lines are the percentage of classification results for which the correct class is within the first $N$ classification outputs as a function of $N$.

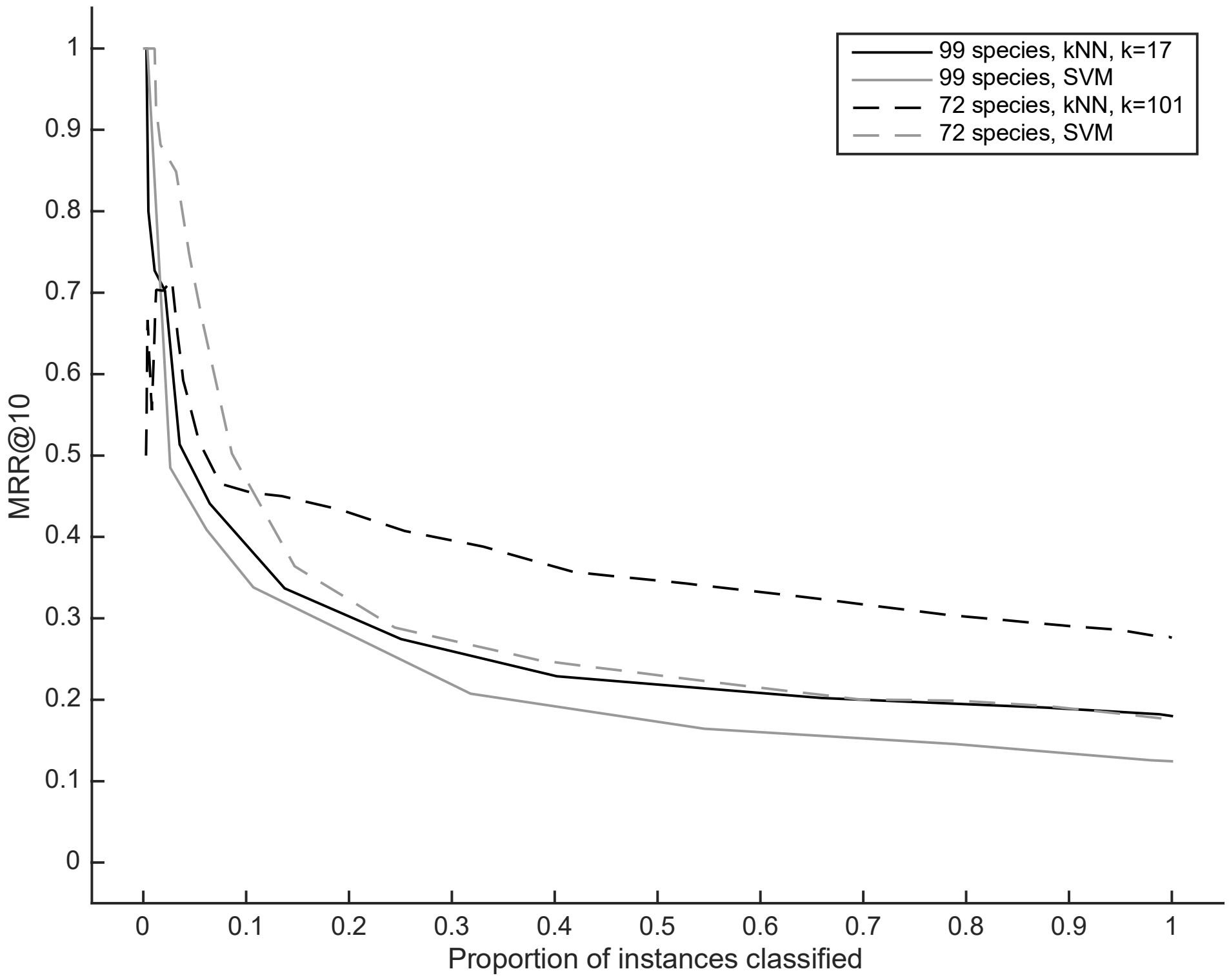

Figure 4: Mean Reciprocal Rank at 10 as a function of the proportion of test instances that are accepted for classification (the experimental setup is the same as in Figure 3).

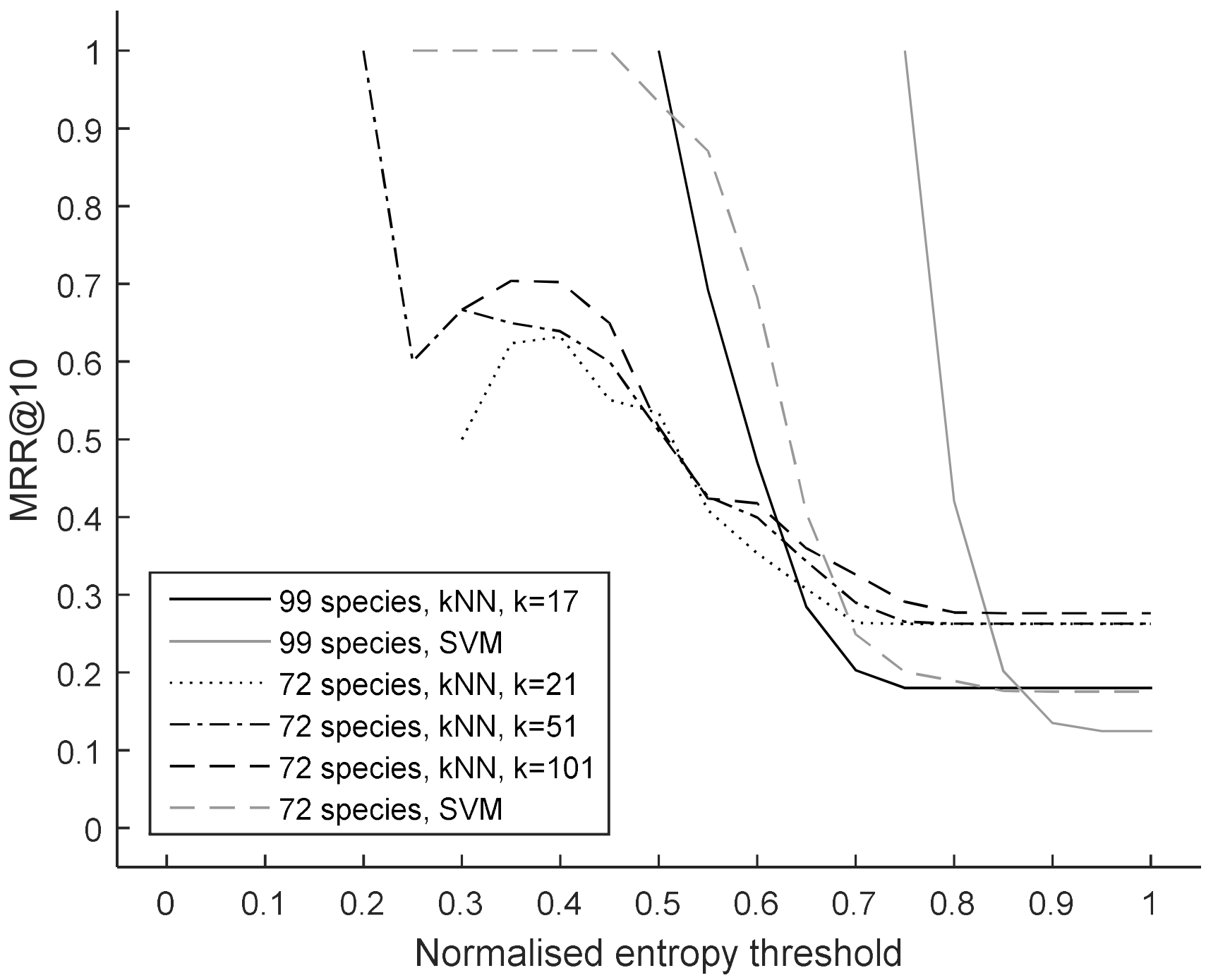

Figure 5: Mean Reciprocal Rank at 10 as a function of the normalised entropy-based threshold (see main text for details of normalisation; the experimental setup is the same as in Figure 4 with two more cases of voting neighbours included in the *k*NN classifier cases)

## Appendix

Audio recordings data and species selection

In Table 2 we list the 99 species that are considered in the paper. In the same table we list the number of training recordings from the xeno-canto (XC) website and testing recordings from the Reference Animal Vocalisations (RAV) collection as well as the total number of available XC training frames per species as selected by the power threshold method (see below and main text). We note that in the species selection we excluded the Eurasian Siskin, which appears in the RAV collection with the binomial name *Carduelis spinus* but in the XC collection as *Spinus spinus* (and because of that was not identified by our download scripts). We also exclude recordings of the Snowy Owl species (*Bubo scandiacus*) for which there are very few recordings in xeno-canto.

The audio recordings that we use are freely accessible at the Animal Sound Archive (www.animalsoundarchive.org/RefSys/Statistics.php) and xeno-canto (www.xeno-canto.org) websites. For the results presented in this paper we used data as downloaded from the websites in March 2015. For the xeno-canto data we used the API provided by that website to query for recordings of the 99 species described above that are rated by users as having the highest quality ('Quality A' rating). We used scripts to parse the html code of each recording's webpage and we screened out recordings that are marked as having other species in the background (we kept only recordings labelled with 'Background none'). In the last section of this appendix we provide identifiers that can be used to retrieve each of these recordings from the online archives.

Table 2: List of 99 species used in the experiments. Species in bold are the 72 species for which there are at least 20000 training frames (see 'Results' section of paper).

| Species | xeno-canto #recordings | xeno-canto #frames | Ref.Anim,Voc. #recordings |
|---|---|---|---|
| **Acrocephalus arundinaceus** | 74 | 103581 | 53 |
| **Acrocephalus palustris** | 126 | 246550 | 21 |
| **Acrocephalus schoenobaenus** | 73 | 145405 | 10 |
| **Acrocephalus scirpaceus** | 80 | 133305 | 28 |
| **Aegolius funereus** | 54 | 99774 | 33 |
| **Alauda arvensis** | 91 | 212959 | 16 |
| Anthus pratensis | 40 | 8124 | 15 |
| **Anthus trivialis** | 75 | 65236 | 32 |
| **Asio otus** | 67 | 67096 | 43 |
| **Athene noctua** | 28 | 22515 | 22 |
| Botaurus stellaris | 10 | 10429 | 22 |
| Branta canadensis | 33 | 17818 | 12 |
| **Buteo buteo** | 24 | 22484 | 14 |
| **Caprimulgus europaeus** | 59 | 198244 | 14 |
| **Carpodacus erythrinus** | 41 | 24154 | 31 |
| **Certhia brachydactyla** | 58 | 33345 | 21 |
| **Certhia familiaris** | 46 | 22469 | 22 |
| **Chloris chloris** | 65 | 70421 | 13 |
| Chlidonias hybrida | 14 | 3335 | 15 |
| **Chroicocephalus ridibundus** | 46 | 92922 | 14 |
| **Corvus cornix** | 49 | 47355 | 10 |
| **Crex crex** | 74 | 114966 | 16 |
| **Cuculus canorus** | 52 | 61179 | 19 |
| **Cyanistes caeruleus** | 128 | 79996 | 13 |
| **Dendrocopos major** | 146 | 84546 | 50 |
| **Dendrocopos medius** | 33 | 29912 | 50 |
| **Dendrocopos minor** | 36 | 32454 | 28 |
| **Dryocopus martius** | 65 | 40447 | 42 |
| Emberiza aureola | 2 | 2110 | 11 |
| **Emberiza calandra** | 80 | 58538 | 45 |
| **Emberiza citrinella** | 107 | 71309 | 83 |
| **Emberiza hortulana** | 42 | 23144 | 279 |
| Emberiza pusilla | 5 | 1830 | 19 |
| Emberiza rustica | 4 | 3057 | 32 |
| **Emberiza schoeniclus** | 67 | 34914 | 77 |
| **Erithacus rubecula** | 185 | 174362 | 26 |
| Ficedula albicollis | 24 | 17589 | 10 |
| **Ficedula hypoleuca** | 60 | 44658 | 39 |
| **Ficedula parva** | 28 | 28816 | 35 |
| **Fringilla coelebs** | 237 | 152401 | 237 |
| Fringilla montifringilla | 19 | 8776 | 18 |
| Fulica atra | 33 | 8144 | 57 |
| Gallinula chloropus | 12 | 3085 | 28 |
| Galerida cristata | 15 | 19039 | 16 |
| **Gallinago gallinago** | 35 | 23951 | 10 |
| **Garrulus glandarius** | 70 | 35501 | 12 |
| **Hippolais icterina** | 54 | 115124 | 11 |
| **Jynx torquilla** | 22 | 36112 | 11 |
| **Locustella fluviatilis** | 53 | 270466 | 13 |
| **Locustella luscinioides** | 24 | 122827 | 12 |
| **Locustella naevia** | 65 | 358506 | 23 |
| **Lophophanes cristatus** | 36 | 27173 | 57 |
| **Loxia curvirostra** | 103 | 67270 | 10 |
| **Lullula arborea** | 34 | 46086 | 13 |
| **Luscinia luscinia** | 61 | 69745 | 127 |
| **Luscinia megarhynchos** | 154 | 190974 | 138 |
| Merops apiaster | 16 | 15281 | 11 |
| Motacilla alba | 52 | 19432 | 14 |
| Muscicapa striata | 29 | 18493 | 10 |
| Nucifraga caryocatactes | 29 | 17914 | 18 |
| **Oriolus oriolus** | 26 | 33629 | 14 |
| **Otus scops** | 34 | 22928 | 16 |
| **Parus major** | 277 | 191591 | 139 |
| **Periparus ater** | 105 | 78444 | 112 |
| **Phoenicurus ochruros** | 44 | 33195 | 17 |
| **Phoenicurus phoenicurus** | 62 | 53842 | 129 |
| Phylloscopus bonelli | 26 | 8033 | 60 |
| Phylloscopus canariensis | 12 | 3966 | 38 |
| **Phylloscopus collybita** | 202 | 107983 | 89 |
| Phylloscopus ibericus | 24 | 17762 | 129 |
| **Phylloscopus sibilatrix** | 66 | 65285 | 29 |
| **Phylloscopus trochilus** | 99 | 73492 | 110 |
| Picus canus | 22 | 11120 | 54 |
| Picus viridis | 40 | 13760 | 23 |
| Podiceps cristatus | 11 | 6503 | 21 |
| Podiceps grisegena | 14 | 9330 | 31 |
| **Poecile montanus** | 68 | 56000 | 16 |
| Porzana parva | 9 | 17682 | 23 |
| Porzana porzana | 21 | 8136 | 66 |
| **Prunella modularis** | 60 | 40650 | 11 |
| **Rallus aquaticus** | 50 | 40067 | 85 |
| Regulus ignicapilla | 31 | 19965 | 14 |
| **Saxicola rubetra** | 43 | 37684 | 54 |
| **Sitta europaea** | 86 | 51781 | 23 |
| **Strix aluco** | 121 | 150023 | 29 |
| **Sylvia atricapilla** | 182 | 265818 | 41 |
| **Sylvia borin** | 70 | 114247 | 26 |
| **Sylvia communis** | 133 | 124219 | 57 |
| **Sylvia curruca** | 87 | 49860 | 26 |
| **Sylvia melanocephala** | 39 | 28953 | 37 |
| **Sylvia nisoria** | 25 | 22166 | 34 |
| Tringa nebularia | 35 | 7901 | 11 |
| **Troglodytes troglodytes** | 140 | 123758 | 35 |
| **Turdus merula** | 188 | 271712 | 118 |
| **Turdus philomelos** | 139 | 300649 | 105 |
| **Turdus pilaris** | 59 | 58383 | 27 |
| **Turdus viscivorus** | 41 | 50322 | 61 |
| **Tyto alba** | 31 | 56740 | 25 |
| **Upupa epops** | 16 | 22670 | 16 |

Spectrogram parameters and power threshold frame selection method

The features we extract are based on FFT-based spectrograms. The spectrogram of each of the training and test recording is computed with a frame length of 20ms, overlap of 50% and rectangular window. The number of FFT points is set to 1024 points for recordings of 48kHz sampling rate and to the first power of two that is higher than this proportion for recordings of other sampling rates (thus corresponding to a minimum distance between consecutive frequency bins of approximately 47Hz). In order to minimize the contribution of low frequency wind noise in the recordings, only the bins corresponding to frequencies between 1kHz and 10kHz are kept. Those frequency limits are within the range of what is typically used in existing bird recognition studies and slight changes around these values should not be expected to have a considerable effect on the system's performance (see e.g. Graciarena et al. 'Acoustic front-end optimization for bird species recognition', IEEE ICAASP, 2010 http://doi.org/10.1109/icassp.2010.5495923).

For the training (XC) recordings we use a power threshold method to select frames from which frame-level features are extracted. This is a modification of the method presented in the works of Briggs et al. (2009) and Stowell & Plumbley (2014a) and it is based on the fact that xeno-canto recordings labelled as 'Quality A' and containing only a single species are generally of high audio quality with practically no other sources of sound than low level ambient noise.

Analysis presented for similar recordings in Stowell & Plumbley (2014a) shows that a power threshold method retaining the 10% most energetic frames achieves high precision i.e. practically all selected frames are bird sound but a possibly high number of bird sound frames is unnecessarily discarded. Here, rather than stipulating a constant ratio of bird to non-bird sound duration across all recordings we used a relative per-recording threshold where frames within a certain power level compared to each recording's maximum level were retained. Namely, for each recording we identify the 1% most energetic frames and we take the mean energy of those frames as an estimate of highest level for that particular recording. Following that, we select all frames of that recording that have power of at least 0.25 of the estimated highest level as corresponding to bird sound and discard the remaining frames. The total numbers of frames per species selected in this way are listed in Table 2.

Frame-level features were then computed (in the case of the training xeno-canto recordings only for the selected frames). Four types of frame-level features were considered, all choices from the various types described in Briggs et al. (2009) and Stowell & Plumbley (2014a). More specifically, after normalising the absolute value of each frame's spectrum to sum unity, we computed the mean (denoted here as $f_{mean}$), standard deviation ($f_{std}$), mode ($f_{mode}$) and the difference between the mode of each selected frame and its successor in the original recording (i.e. regardless of whether the subsequent frame was selected or not in the power threshold procedure, denoted as $\Delta f_{mode}$). The first two of these frame-level features ($f_{mean}$ and $f_{std}$) closely resemble those described in Briggs et al. (2009) while the last two ($f_{mode}$) and $\Delta f_{mode}$) those described in Stowell & Plumbley (2014a)

For the training (XC) data, the individual sequences of frame-level features from each recording of the same species are lined-up in one sequence (which preserves the original frame sequence in each audio recording but contains 'cuts' due to rejected frames and due to the line-up of the different recordings) and sequences of features obtained from 100 frames are taken. Hence, each such 100-frame instance generally contains frame-level features from both consecutive and non-consecutive frames and from both the same and different recordings. Binned histogram and time-summarisation features for the training (XC) recordings are computed on the basis of such 100-frame sequences. We consider three types of binned histograms: (i) two-dimensional histograms of the $f_{mean}$ and $f_{std}$ frame-level features with 100 and 50 bins spaced uniformly from 1kHz to 10kHz along the two respective dimensions (corresponding to 5000-dimensional feature vectors), (ii) one-dimensional histograms of the $f_{mode}$ feature with 100 bins spaced uniformly from 1kHz to 10kHz along the $f_{mode}$ dimension (corresponding to 100-dimensional feature vectors) and (iii) two-dimensional histograms of the $f_{mode}$ and $\Delta f_{mode}$ pair of features with 100 bins spaced uniformly from 1kHz to 10kHz along the $f_{mode}$ dimension and 50 bins spaced uniformly from -2kHz to 2kHz along the $\Delta f_{mode}$ dimension (corresponding to 5000-dimensional feature vectors). We use the time-summarisation features found to give the best performance in Stowell & Plumbley (2014a), namely 6-dimensional feature vectors comprising the 5$^{th}$, 50$^{th}$ and 95$^{th}$ percentiles of the $f_{mode}$ and the 50$^{th}$, 75$^{th}$ and 95$^{th}$ percentiles of the $\Delta f_{mode}$ frame-based features. Standardization is applied to the training data prior to training the SVM classifier and the same linear transformation is applied to the test data prior to the prediction stage.

The test (RAV) recordings are used as they are with no frame selection applied and one binned histogram or time-summarisation feature vector is computed on the basis of each whole (RAV) recording.

Tie-breaking bias in the kNN voting score

The use of the AUC in the case of the $k$NN classifier becomes problematic for low values of the $k$ parameter (number of voting neighbours) as the returned probability estimate scores are very coarsely quantized. This results in a misleading increase in the AUC score as the $k$ parameter is increased. Furthermore, in order to obtain an ordered list of most likely species, a tie breaking rule needs to be introduced in cases where two classes (species) have the same number of voting neighbours. We did that by adding an adjustment bias to the list of $k$NN voting scores (class membership probabilities) for each test instance. This bias was a uniformly spaced grid of length equal to the number of classes, ordered by the class that contained the closest training instance to the given test instance and ranging from 0 up to $1/(k*M)$ for $M$ an arbitrarily chosen number greater than unity. It can be readily seen that, with $M>1$, the maximum value of this bias is smaller than the smallest possible difference between class membership probabilities (which is $1/k$, the reciprocal of the number of voting neighbours). Hence the addition of this bias has the effect of ordering classes with equal number of voting neighbours without altering the overall voting score order. It should be noted that with this bias addition, the voting scores do not sum up to unity. However this does not affect the shape of the AUC curve.

SVM classifier implementation

The Matlab interface of the LIBSVM software package with its default settings for multiclass probabilistic classification is used for the results presented here (Chang & Lin 2011). These settings amount to a radial basis function kernel and a "one-against-one" method for recasting the problem of training a k-class multiclass classifier to the training of k(k-1)/2 binary SVM classifiers. Furthermore, LIBSVM provides an estimate of the vector of posterior class-membership probabilities given the observed test instance by using the method described in (Wu et al. 2004) for the transformation of the one-against-one SVM class decision functions to pairwise class posterior probabilities and consequently the method described in (Lin et al. 2007) for the estimation of posterior probabilities in a multiclass setup.

Description of the provided Matlab data

The Matlab cell (99 rows and 5 columns) stored in the variable 'upload_cell' inside the 'upload_data.mat' Matlab file contains:

1. The Latin names of the 99 species in its first column
2. Numeric identifiers for the xeno-canto training recordings in its second column. For example, the first recording of the first species ('Acrocephalus arundinaceus') can be retrieved by

   >> upload_cell{1,2}(1)

   ans =

247482

and the corresponding xeno-canto recording can be accessed online as

http://www.xeno-canto.org/247482

3. Text identifiers for the Animal Sound Archive test recordings in its third column. For example, the first test recording for the first species ('Acrocephalus arundinaceus') can be retrieved by

```
>> upload_cell{1,3}(1)

ans =

   'AcrAru00001'
```

and the corresponding the Animal Sound Archive recording can be accessed as 'AcrAru00001.wav' in the zip file downloaded via the 'Acrocephalus arundinaceus' web-link in http://www.animalsoundarchive.org/RefSys/Statistics.php

4. Text identifiers for the Animal Sound Archive test recordings used in the balanced test dataset (10 recordings per species) in its fourth column.

5. A boolean flag signifying if the corresponding species is in the 72 species dataset in its fifth column